\documentclass[preprint,showpacs,preprintnumbers,amsmath,amssymb]{revtex4}

\usepackage{graphicx}
\usepackage{dcolumn}
\usepackage{bm}

\begin{document}

%
%

\title{Microwave magnetoplasmon absorption by a 2DEG stripe. }

\author{O. M.~Fedorych\footnote{Present address: Technische Universität Dortmund Experimentelle Physik II, Dortmund, Germany.}}
\address{Grenoble High Magnetic Field Laboratory, CNRS, Grenoble, France}
\author{S.~Moreau}
  \address{Grenoble High Magnetic Field Laboratory, CNRS,
  Grenoble, France}
\author{M.~Potemski}
  \address{Grenoble High Magnetic Field Laboratory, CNRS,
  Grenoble, France}
\author{S. A.~Studenikin\footnote{sergei.studenikin@nrc.ca}}
\address{Institute for Microstructural Sciences, NRC, Ottawa,
  Ontario, Canada}
\author{T. Saku}
\address{ NTT Basic Research Laboratories, NTT Corporation, Atsugi, Japan}
\author{Y. Hirayama}
\address{ NTT Basic Research Laboratories, NTT Corporation, Atsugi, Japan}

\begin{abstract}
Microwave (MW) absorption by a high mobility 2DEG has been
investigated experimentally using sensitive Electron Paramagnetic Resonance (EPR) cavity technique.
It is found that MW absorption spectra are chiefly governed by confined magnetoplasmon excitations in a 2DEG stripe.
Spectra of the 2D magnetoplasmons are studied as a function of magnetic field,  MW frequency and carrier density.
The electron concentration is tuned by illumination and monitored using optical photoluminescence technique.
\end{abstract}

\maketitle

\section{Introduction}

 Currently there is an increased interest in microwave (MW) properties of high-mobility two-dimensional electron gas (2DEG), partially, due to recent discovery of microwave induced resistance oscillations (MIROs) which under certain conditions may evolve into so called zero resistance states.\cite{Mani,Zudov} The microscopic physics of MIROs is still under debate.  This phenomenon is observed in \emph{dc} transport measurements and first attempts to observe MIRO-like effects in absorption/reflection experiments revealed very different behavior as compared to \emph{dc} transport.\cite{IEEE}
This work is devoted to in-detail investigation of the MW absorption on a high mobility 2DEG stripe by using a sensitive EPR resonator technique. Several peaks are observed in the absorption spectra of the rectangular 2DEG sample. In order to confirm that these peaks originate from confined magneto-plasmon resonances, a careful study of the peaks is performed as a function of magnetic field, microwave frequency and electron concentration.

\section{Experiment}
Two-dimensional electron gas was confined in a 20 nm wide GaAs/AlGaAs quantum well, similar to the one studied in ref.\cite{Nature}.
 A rectangular  stipe was cleaved from this sample to the dimensions of approximately 0.8$\times$2$mm^2$.
 The electron density in dark was $n_{2DEG}=2.0\times 10^{11}cm^{-2}$, and mobility was $\simeq 3\times 10^6 cm^2/Vs$. After full illumination the electron concentration decreased to $1.3\times 10^{11}cm^{-2}$.
 An optical fiber was used to deliver He-Ne laser radiation into the cavity through a small hall in the plunger.  Varying the illumination intensity, we were able to change 2DEG concentration gradually and simultaneously determine it from the analysis of the magneto-photoluminescence spectra.\cite{Moreau,Moreau2}
\begin{figure}
\includegraphics[width=5in]{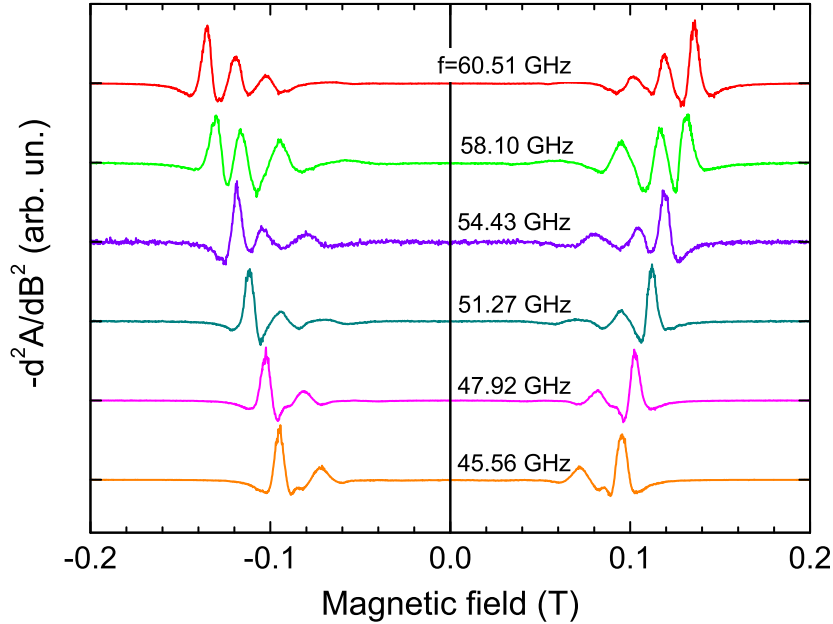}
\caption{Experimental traces of the negative second derivative of the MW power absorption as a function of magnetic field at different MW frequencies, T=$4$~$K$.}
 \label{fig1}
\end{figure}

For the MW absorption measurements we employ an Electron Paramagnetic
Resonance (EPR) spectrometer with a cylindric cavity tuned to the $TE011$
mode. The electric field component for this mode has curling geometry and lays in the $xy$-plane. The  electric field has two  zeros - in the center and at the cavity edge, and it reaches the maximum value at the middle of the cavity radius.
In $z$ direction along the cavity height the electric field reaches maximum at the medium of the cavity height. A movable cylindrical plunger is used to tune the cavity resonance frequency in the range between 40 and 60 $GHz$ by
adjusting the cavity height between 3 and 8 $mm$.  Modulation of the
external magnetic field in combination with a lock-in detection
technique allows us to improve sensitivity to small signals and
increase signal to noise ratio. A detailed description of the EPR
spectrometer setup can be found in Refs.\cite{Seck,Seck2}.

The rectangular 2DEG sample is placed at the bottom of the
cylindric cavity along the radius so that the external magnetic
field is normal to the 2DEG plane, and the MW electric field component of $TE011$ mode lays in the 2DEG plane  and is directed along the short side of the rectangular. This geometry is preferable for excitation of confined
plasmons across the bar.
In this work we position our sample in the "face down" configuration in which the active 2DEG layer is in in close proximity to the plunger plate and, therefore, the coupling of conduction electrons to the MW cavity is minimized and produces minimum distortion of the cavity resonance.

Figure~1 shows experimental traces of the second derivative of the microwave absorption in respect to the magnetic field for both direction of magnetic field.  This experiment is performed at liquid helium temperature for different MW frequencies. It is known that the maximum positions of the second derivative (with minus sign) coincide with the maxima of the original function, but peaks become sharper and clearer for the analysis.\cite{Stu1986}
Several peaks are evident in the
traces in Fig.~1, which number increases with frequency.  The position and shape of the peaks are symmetric relative to zero magnetic field indicating the correct tuning of the setup. The magnetic field and frequency dependencies of the peak position are characteristic for the confined magnetoplasmons  in  finite size 2DEG samples.\cite{Stern1967}
\begin{figure}
\includegraphics[width=5in]{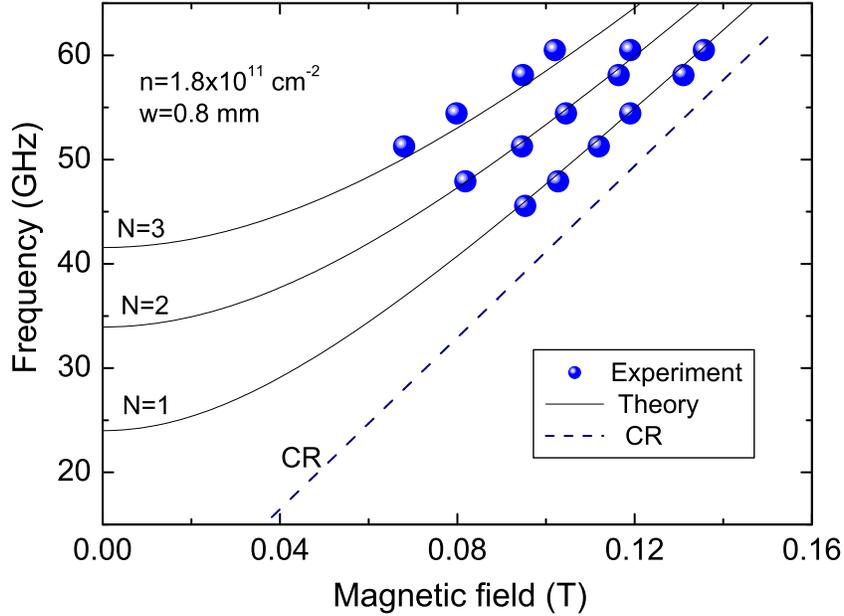}
\caption{Frequency position of the resonance peaks \emph{vs.} magnetic field (points). Solid lines are the theoretical dependencies according to Eq.~(1), electron concentration $n_{s}=1.8\times10^{11}$  $cm^{-2}$. The dash line corresponds to the cyclotron
resonance  calculated for effective electron mass in GaAs $m^*=0.068 m_0$, where $m_0$ is the mass of free electron. }
 \label{fig2}
\end{figure}

\section{Discussion}
The dispersion of confined plasmons in zero magnetic field in quasi-electrostatic approximation is described by the following equation: \cite{Stern1967}
\begin{equation}\label{plasma}
    \omega_{p}^{2}(q,n_{s})= \frac{2 \pi e^{2}n_{s}}{m^{\star}\varepsilon}q,
\end{equation}
where $n_{s}$ is the sheet concentration of 2D carriers, and
$\varepsilon$ is the  effective dielectric permittivity of the
surrounding medium. In our case of long wavelength plasmons ($\lambda_p \sim$ 1 \emph{mm} with $\lambda \gg d$, where \emph{d} is the depth of 2DEG below the surface) we used $\varepsilon=(\varepsilon_{GaAs}+1)/2$ with $\varepsilon_{GaAs}=12.8$ being a dielectric constant of gallium arsenide.\cite{Kukushkin,Govorov}

In our case, when the retardation effects are not important\cite{MikhailovPRB05}, the 2D plasmon frequencies in magnetic field are described by the following equation:
\begin{equation}\label{2}
\omega^{2}(q,n_{s},B)=\omega^{2}_{c}(B)+\omega_{p}^{2}(q,n_{s}),
\end{equation}
where $\omega_{p}$ is plasmon cyclic frequency at zero magnetic field, $q=N\pi/\textrm{w}$ denotes the plasmon wave vector; N=1,2,3... is the plasmon mode, w is the sample width, $\omega_c = eB/m^*$ is the cyclotron frequency, $m^*$ is the electron effective mass. In our calculations we use $m^*/m_0$=0.068 estimated for this particular quantum well sample.\cite{Moreau,Moreau2}
The peak positions of the magnetoplasmon resonances are plotted in Fig.~2 as a function of magnetic field. It is evident that Eq.(2) describes the experimental data on  this sample quite well, suggesting that the observed peaks are indeed due to the confined magnetoplasmon resonances.
\begin{figure}
\includegraphics[width=5in]{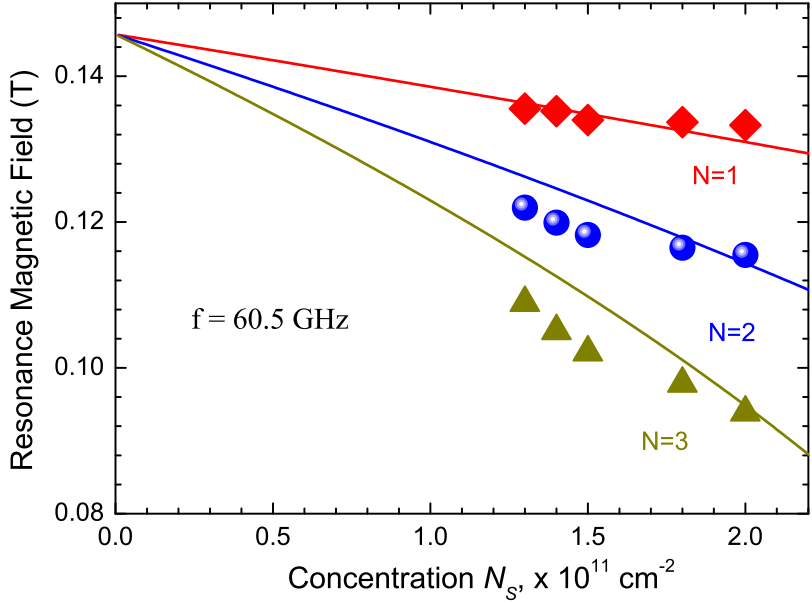}
\caption{The magnetic field position of magnetoplasmon resonance peaks (solid points) for $f_{MW}$=60~GHz as a function of electron density. Electron density was tuned by illumination intensity. Solid lines are calculated theoretical dependencies according to Eq.~(3).}
\label{fig3}
\end{figure}

To obtain further evidences that the observed peaks in Fig.~1 are due
to plasmon resonances, we have also studied  their dependence as a function of the electron density.  As it is mentioned above, the free carrier density is tuned by varying the illumination intensity.
As it follows from Eqs. (1) and (2), the resonance field position of the magnetoplasmon peaks \emph{vs.} electron concentration is described by the following equation:
\begin{equation}\label{3}
B_{N}=\frac{m^*}{e}\sqrt{f^{2}_{0}-\frac{n_s e^2}{2 m^*}\frac{N}{\varepsilon \textrm{w}}},
\end{equation}
where $f_{0}$ is the MW frequency at which the  measurements are performed, \emph{N} is the plasmon mode.  A useful practical equation to calculate plasmon peak position is given by the following formulae:
\begin{equation}\label{4}
B_{N}=\frac{1}{411.7}\sqrt{f^{2}_{0}-186.2\frac{n_s N}{\varepsilon \textrm{w}}},
\end{equation}
where $B_N$ is measured in Tesla, $f_0$ in GHz, $n_s$ is electron surface density in $10^{11} cm^{-2}$, and w is the stripe width in \emph{cm}.

The results of this study \emph{vs.} electron concentration are presented in Fig.~3. It is seen from the figure that in accordance with Eq.~3 the splitting between magneto-plasmon modes increases with the density. It is clear that agrement between the experiment (points) and  the theory (solid lines) is fairly good, confirming again that the observed structure  is due to confined magnetoplasmons.
\newpage
\section{Conclusions}
In summary, the MW absorption by a 2DEG stripe made of a fairly high-mobility GaAs/AlGaAs wafer was investigated using a sensitive EPR cavity technique. It was shown that the magnetoplasmons play a dominant role in microwave experiments on 2DEG samples. The 2D plasmon spectra were
studied as a function magnetic field, microwave frequency, and carrier concentration.  Experimental results were compared with the theoretical predictions and a good agreement between both was found.

\end{document}